\documentclass[%
reprint,
superscriptaddress,
nofootinbib,
amsmath,amssymb,
aps,
prl]{revtex4-2}
\usepackage{graphicx}
\usepackage{dcolumn}
\usepackage{bm}
\usepackage{mathtools}
\usepackage{color}
\usepackage[colorlinks=true,linkcolor=blue,citecolor=blue]{hyperref}

\renewcommand{\vec}[1]{\boldsymbol{#1}}
\newcommand{\bra}{\langle}
\newcommand{\ket}{\rangle}

\begin{document}

\title{Impact of two-body currents on magnetic dipole moments of nuclei}

\author{T.~Miyagi}
\email{miyagi@theorie.ikp.physik.tu-darmstadt.de}
\altaffiliation[Present address: ]{Center for Computational Sciences, University of Tsukuba, 1-1-1 Tennodai, Tsukuba 305-8577, Japan} %
\affiliation{Technische Universit\"at Darmstadt, Department of Physics, 64289 Darmstadt, Germany}
\affiliation{ExtreMe Matter Institute EMMI, GSI Helmholtzzentrum f\"ur Schwerionenforschung GmbH, 64291 Darmstadt, Germany}
\affiliation{Max-Planck-Institut f\"ur Kernphysik, Saupfercheckweg 1, 69117 Heidelberg, Germany}
\author{X.~Cao}
\email{xuchenc2@illinois.edu}
\affiliation{Department of Physics and Institute for Condensed Matter Theory, University of Illinois at Urbana-Champaign, 1110 West Green Street, Urbana, Illinois 61801-3080, USA}
\author{R.~Seutin}%
\email{rseutin@theorie.ikp.physik.tu-darmstadt.de}
\affiliation{Max-Planck-Institut f\"ur Kernphysik, Saupfercheckweg 1, 69117 Heidelberg, Germany}
\affiliation{Technische Universit\"at Darmstadt, Department of Physics, 64289 Darmstadt, Germany}
\affiliation{ExtreMe Matter Institute EMMI, GSI Helmholtzzentrum f\"ur Schwerionenforschung GmbH, 64291 Darmstadt, Germany}
\author{S.~Bacca}
\email{s.bacca@uni-mainz.de}
\affiliation{Institute of Nuclear Physics, Johannes Gutenberg-Universit\"at Mainz, 55099 Mainz, Germany}
\affiliation{PRISMA+ Cluster of Excellence, Johannes Gutenberg-Universit\"at Mainz, 55099 Mainz, Germany}
\author{R~.F.~Garcia~Ruiz}
\email{rgarciar@mit.edu}
\affiliation{Department of Physics, Massachusetts Institute of Technology, Cambridge, MA 02139, USA}
\author{K.~Hebeler}
\email{kai.hebeler@physik.tu-darmstadt.de}
\affiliation{Technische Universit\"at Darmstadt, Department of Physics, 64289 Darmstadt, Germany}
\affiliation{ExtreMe Matter Institute EMMI, GSI Helmholtzzentrum f\"ur Schwerionenforschung GmbH, 64291 Darmstadt, Germany}
\affiliation{Max-Planck-Institut f\"ur Kernphysik, Saupfercheckweg 1, 69117 Heidelberg, Germany}
\author{J.~D.~Holt}
\email{jholt@triumf.ca}
\affiliation{TRIUMF, 4004 Wesbrook Mall, Vancouver BC V6T 2A3, Canada}
\affiliation{Department of Physics, McGill University, Montréal, QC H3A 2T8, Canada}
\author{A.~Schwenk}
\email{schwenk@physik.tu-darmstadt.de}
\affiliation{Technische Universit\"at Darmstadt, Department of Physics, 64289 Darmstadt, Germany}
\affiliation{ExtreMe Matter Institute EMMI, GSI Helmholtzzentrum f\"ur Schwerionenforschung GmbH, 64291 Darmstadt, Germany}
\affiliation{Max-Planck-Institut f\"ur Kernphysik, Saupfercheckweg 1, 69117 Heidelberg, Germany}

\begin{abstract}
We investigate the effects of two-body currents on magnetic dipole moments of medium-mass and heavy nuclei using the valence-space in-medium similarity renormalization group with chiral effective field theory interactions and currents.
Focusing on near doubly magic nuclei from oxygen to bismuth, we have found that the leading two-body currents globally improve the agreement with experimental magnetic moments. Moreover, our results show the importance of multi-shell effects for $^{41}$Ca, which suggest that the $Z=N=20$ gap in $^{40}$Ca is not as robust as in $^{48}$Ca.
The increasing contribution of two-body currents in heavier systems is explained by the operator structure of the center-of-mass dependent Sachs term.
\end{abstract}

\maketitle

Nuclear magnetic dipole moments are a key probe to explore the structure of atomic nuclei.
For odd-mass systems, the simplest description of magnetic moments is to consider only the contribution from the last unpaired nucleon, known as the single-particle or Schmidt limit~\cite{Schmidt1937}.
An experimental deviation from the Schmidt limit indicates the impact of many-body contributions to the magnetic moment, with important contributions from core-polarization effects~\cite{Arima1987,Towner1987,Sassarini2022}. 
Since the magnetic moments are sensitive to shell structure, they provide an important probe of nuclear structure and shell closures, complementary to high $2^+$ excitation energies, high separation energies, and more inert radii at magic numbers.
Recent experimental efforts have thus focused on the evolution of magnetic moments along isotopic chains~\cite{Yang2023}.
From the theoretical side, providing an accurate description of magnetic moments in medium-mass and heavy nuclei has been a major challenge.
The comparison with experiments often requires the use of adjustable parameters that are commonly fitted to improve agreement with experimental data in specific regions of the nuclear chart (see, e.g., Ref.~\cite{Yang2023}).
For a reliable description of magnetic dipole moments, it is important to perform controlled nuclear structure calculations with many-body electromagnetic (EM) operators. The goal of this work is a first global ab initio survey of magnetic moments near doubly magic nuclei from oxygen to bismuth.

In the past decades, great progress has been made in advancing ab initio calculations to medium-mass and heavy nuclei~\cite{Morris2018,Arthuis2020,Hergert2020,Hebeler2020,Stroberg2021,Miyagi2022}, culminating in the recent ab initio calculation of $^{208}$Pb~\cite{Hu2022}.
At the same time, ab initio calculations have explored EM observables and weak transitions including contributions beyond the standard one-body operators~\cite{Pastore2009,Gazit2009,Marcucci2012,Piarulli2013,Pastore2013,Bacca2014,Gysbers2019,King2020,Friman-Gayer2021}. However, these efforts have so far focused on light nuclei, except for a global study of beta decays of medium-mass nuclei up to $^{100}$Sn~\cite{Gysbers2019}. The latter work showed that many-body correlations and two-body currents (2BC) are key to explain the quenching puzzle of beta decays. Here, we focus on magnetic moments up to bismuth, including both many-body correlations and for the first time the leading EM 2BC.

Another motivation for this work is the recent precision measurements of the magnetic dipole moments of indium isotopes~\cite{Vernon2022}. The experimental results showed a striking jump at $N=82$ towards the Schmidt limit, supporting the expected magic number at $N=82$.
However, the size of the magnetic moments could not be reproduced by ab initio calculations using the valence-space in-medium similarity renormalization group (VS-IMSRG) approach with only one-body EM operators~\cite{Vernon2022}.
Similar deficiencies were also seen in ab initio calculations of medium-mass nuclei (see, e.g., Refs.~\cite{Klose2019,Heylen2021,Powel2022}).
As the VS-IMSRG approach takes into account many-body correlations, such as core-polarization effects, in a non-perturbative way, these deficiencies have been attributed to the neglected higher-order two-body contributions to EM operators from pion-exchange currents as well as shorter-range contributions. For light nuclei ($A < 20$), the significance of 2BC for magnetic moments and electromagnetic transitions has been shown in quantum Monte-Carlo and no-core shell model calculations (see, e.g., Refs.~\cite{Piarulli2013,Pastore2013,Bacca2014,Martin2023}).
In this Letter, we provide a global survey of the impact of the leading 2BC for magnetic dipole moments from medium-mass to heavy nuclei using the ab initio VS-IMSRG.
Since the vector currents also enter precision calculations of weak decays~\cite{Brodeur2023}, testing 2BC against EM observables is a critical test of electroweak operators for applications to fundamental symmetry tests with rare decays.

Chiral effective field theory (EFT) is a low-energy expansion of quantum chromodynamics with nucleons and pions as degrees of freedom. It provides a systematic expansion of nuclear forces~\cite{Epelbaum2009,Machleidt2011} and consistent electroweak currents~\cite{Krebs2020}. In the EM sector, one- and two-body currents have been derived up to next-to-next-to-next-to-leading order (N$^{3}$LO)~\cite{Pastore2009,Pastore2013,Kolling2009,Kolling2011}. Here, we focus on the magnetic dipole operator, which is defined as $\vec{\mu} = -i \lim_{q\to 0} \nabla_{\vec{q}} \times \vec{j}(\vec{q})/2$ with the EM spatial current $\vec{j}(\vec{q})$, where $\vec{q}$ is the momentum transfer carried by photon. In the following, we will consider the magnetic moment in the $z$-direction. In many-body calculations, usually only the magnetic dipole operator at the one-body level, $\mu_{\rm 1B}$, is used. This takes the well known form:
\begin{equation}
\mu_{\rm 1B} =  \mu_{N} \sum_{i} \left(g^{l}_{i} l_{i,z} + g^{s}_{i} \sigma_{i,z}\right) \,,
\end{equation}
with the magneton of the proton, $\mu_{N} = \frac{e\hbar}{2m_{p}}$ with unit charge $e$ and proton mass $m_{p}$, and $l_{i,z}$ and $\sigma_{i,z}$ are the $z$-component of orbital angular momentum and spin operators for $i$-th nucleon. $g^{l}_{i}$ and $g^{s}_{i}$ are the orbital and spin $g$-factor, with $g^{l}_{\rm proton}=1$, $g^{l}_{\rm neutron}=0$, $g^{s}_{\rm proton}=2.792$, and $g^{s}_{\rm neutron} = -1.913$~\cite{CODATA2018}.
In this work, we consider the leading 2BC, $\mu_{\rm 2B}$, given by the parameter-free pion-exchange contributions~\cite{Krebs2020}. In coordinate space, these are given by the intrinsic and Sachs terms:
\begin{align}
\label{eq:mu_2BC}
\mu_{\rm 2B} &= \sum_{i<j}\mu^{\rm intr}_{ij} + \mu^{\rm Sachs}_{ij} \,, \\
\label{eq:mu_intr}
\mu^{\rm intr}_{ij} &= \mu_{N} (\vec{\tau}_{i}\times \vec{\tau}_{j})_{z}  \vec{V}_{{\rm intr},z}(\vec{r}_{ij}) \,, \\
\label{eq:mu_Sachs}
\mu^{\rm Sachs}_{ij} &= \mu_{N} (\vec{\tau}_{i}\times \vec{\tau}_{j})_{z} 
(\vec{R}_{ij} \times \vec{r}_{ij} )_{z} V_{\rm Sachs}(r_{ij}) \,.
\end{align}
Here, $\vec{r}_{ij}\equiv \vec{r}_{i}-\vec{r}_{j}$ and $\vec{R}_{ij}\equiv (\vec{r}_{i}+\vec{r}_{j})/2$ are the relative and center-of-mass coordinates of the nucleons $i$ and $j$, respectively. $\vec{\tau}_{i}$ is the isospin operator of the $i$-the nucleon, and the detailed expressions for $\vec{V}_{\rm intr}(r_{ij})$ and $V_{\rm Sachs}(r_{ij})$ are, e.g., given in Ref.~\cite{Seutin2023}. The numerical implementation of $\mu_{\rm 2B}$ is also provided in the \texttt{NuHamil} code~\cite{Miyagi2023} used in this work.

\begin{figure}[t]
    \centering
    \includegraphics[clip,width=\columnwidth]{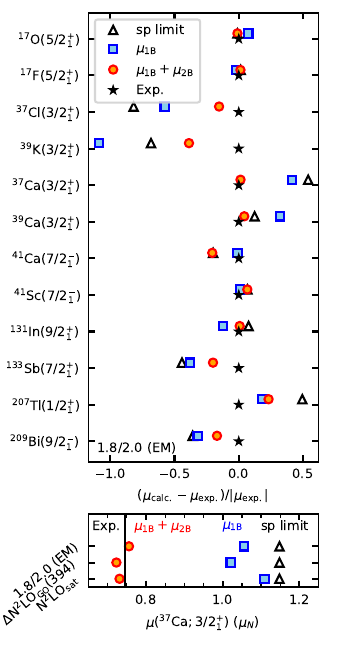}
    \caption{Magnetic dipole moments of near doubly magic nuclei from $A=17-209$ computed with the 1.8/2.0 (EM) interaction relative to the experimental values (top) and of $^{37}$Ca with the 1.8/2.0 (EM), $\Delta{\rm N}^{2}{\rm LO}_{\rm GO}(394)$~\cite{Jiang2020}, and ${\rm N}^{2}{\rm LO}_{\rm sat}$~\cite{Ekstrom2015} interactions (bottom). 
    Results are shown at the one-body level, $\mu_{\rm 1B}$ (blue squares), and including 2BC, $\mu_{\rm 1B}+\mu_{\rm 2B}$ (red circles) obtained with the {\it ab initio} VS-IMSRG(2). The experimental dipole moments (stars) are taken from Ref.~\cite{Stone2005,Vernon2022}. In addition, we show  the simple single-particle (sp) limit (without many-body correlations and without 2BC).}
    \label{fig:mu_1p_1h}
\end{figure}

In this work, we employ the VS-IMSRG~\cite{Tsukiyama2012,Bogner2014,Hergert2016,Stroberg2017,Stroberg2019} to compute the magnetic dipole moments for various near doubly magic nuclei.
The VS-IMSRG calculation starts from nucleon-nucleon (NN) plus three-nucleon (3N) interactions based on chiral EFT, which are expressed in spherical harmonics-oscillator basis states. 
We use the 1.8/2.0 (EM) interaction~\cite{Entem2003,Hebeler2011}, which is fitted to NN scattering phase shifts, the $^{3}$H binding energy, and the $^{4}$He charge radius.
This interaction can reproduce the experimental ground-state energies up to $A\sim 100$~\cite{Simonis2017,Morris2018,Stroberg2021,Miyagi2022,Hebeler2023}, while it provides somewhat too small radii.
The Hamiltonian is first normal ordered with respect to the ensemble reference state~\cite{Stroberg2017}.
Then, we construct an approximate unitary transformation~\cite{Morris2015} with the VS-IMSRG at the normal-ordered two-body level to decouple a selected valence space from the remaining many-body configurations, referred to as the VS-IMSRG(2).
With the same transformation, the $\mu_{\rm 1B}$ and $\mu_{\rm 2B}$ operators are evolved consistently with the Hamiltonian~\cite{Parzuchowski2017a}.
Note that a relaxation of the two-body approximation is important to quantify the many-body uncertainties of the VS-IMSRG(2). This has been achieved recently for the Hamiltonian~\cite{Heinz2021}, but is an ongoing development for general operators.
The calculational setup and convergence for the heaviest nucleus studied, $^{209}$Bi, is summarized and discussed in Supplemental Material.
This demonstrates that the magnetic dipole moments are well converged in terms of the many-body basis space.
The ground-state energies for the heaviest systems, $^{207}$Tl and $^{209}$Bi, are $-1660 \pm 19$\,MeV and $-1671 \pm 19$\,MeV, respectively, after the model-space extrapolation based on Refs.~\cite{Furnstahl2015,Miyagi2022} and adopting a 3\% error of the correlation energy from the VS-IMSRG(2) approximation~\cite{Heinz2021}.
Compared to experiment, the employed 1.8/2.0 (EM) interaction provides slight overbinding in the $A\sim 200$ region, which is consistent with the overbinding found in infinite matter calculations~\cite{Hebeler2011}.
Finally, we also checked that the $\mu_{\rm 2B}$ contribution is decreased by $\lesssim 10$\% in $^{37}$Ca through momentum-space regulators, which is a small effect for the total magnetic moment. We have thus used the unregularized coordinate-space $\mu_{\rm 2B}$ operator in this work.

The top panel in Fig.~\ref{fig:mu_1p_1h} shows our results for the magnetic dipole moments of near doubly magic nuclei from $A=17-209$ computed with the VS-IMSRG(2) relative to the experimental values. The simple single-particle limit is a reasonable starting point for all cases, but cannot explain experiment due to the neglected contributions from many-body correlations and 2BC. Our VS-IMSRG calculation at the one-body level, $\mu_{\rm 1B}$, provide an improvement in several cases due to the inclusion of many-body correlations, primarily from core-polarization effects. However, only after 2BC are included, with  $\mu_{\rm 1B}+\mu_{\rm 2B}$, we find an overall significantly improved agreement with experiment. The improved agreement is present in all cases, except for a small 2BC effect in $^{207}$Tl and a small deterioration in $^{41}$Ca, which can be explained by multi-shell effects, as discussed next. 
We observed that the improvement does not depend on the employed interaction. In the bottom panel of Fig.~\ref{fig:mu_1p_1h}, the magnetic dipole moment for the ground state of $^{37}$Ca is shown for other established NN+3N interactions~\cite{Jiang2020,Ekstrom2015}.
We note that the 2BC contributions are always positive (negative) in the odd $Z$ ($N$) systems studied here, reflecting the isovector nature of the intrinsic and Sachs terms, Eqs.~\eqref{eq:mu_intr} and~\eqref{eq:mu_Sachs}.

\begin{figure}[t]
    \centering
    \includegraphics[clip,width=\columnwidth]{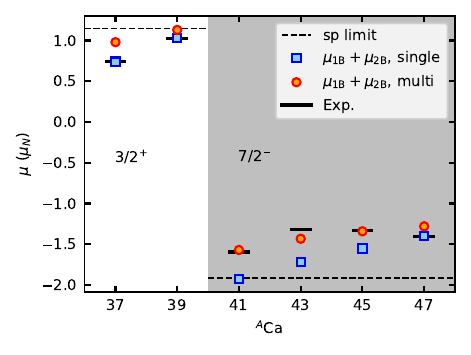}
    \caption{Magnetic dipole moments of the odd-mass calcium isotopes computed with the VS-IMSRG(2) including 2BC, in comparison to experiment~\cite{Ruiz2015}. In addition to the single major-shell results (shown for $^{41}$Ca in Fig.~\ref{fig:mu_1p_1h}) we present results for the multi-shell valence-space calculations.}
    \label{fig:mu_Ca}
\end{figure}

A possible reason for the deterioration with 2BC in $^{41}$Ca is that excitations across the $N=Z=20$ shell closure are not fully taken into account in the $pf$-shell valence space.
As shown in Ref.~\cite{Caurier2001}, the behavior of charge radii in the calcium isotopes suggests that the $^{40}$Ca core is not as robust as $^{48}$Ca.
In Ref.~\cite{Miyagi2020}, we also observed particle-hole excitations across the $N=Z=20$ gap. 
To include these excitation effects explicitly, we perform the calculations with a multi-shell valence space above the $^{28}$Si core (see Ref.~\cite{Miyagi2020} for details). Figure~\ref{fig:mu_Ca} shows the magnetic dipole moments with 2BC for the calcium isotopes for the calculations based on a single major-shell and a multi-shell valence space.
The single major-shell results are for the $sd$ shell for $A < 40$ and $pf$ shell for $A > 40$. They show reasonable agreement with experiment especially in $A < 40$ while they significantly underestimate experiment except for $^{47}$Ca, which is expected from the doubly magic nature of $^{48}$Ca.
As we extend the valence space to capture the excitations across $N=Z=20$, the multi-shell results in $A>40$ are greatly improved.
This shows that the reproduction of experiment at the one-body level for $^{41}$Ca, 
$\mu_{\rm 1B}$ computed in the single major shell in Fig.~\ref{fig:mu_1p_1h}, is accidental, and the $^{40}$Ca core is broken. Moreover, our results in Fig.~\ref{fig:mu_Ca} show that the shell closures of the $^{48}$Ca are more robust than $^{40}$Ca, since both calculations nearly coincide in $^{47}$Ca.
Finally, we have checked that this effect is small for $^{41}$Sc, where the single-shell calculation agrees well with experiment, as shown in Fig.~\ref{fig:mu_1p_1h}. In this case, the multi-shell calculation changes the magnetic moment only by 6\%. Note however that this does not mean that the single- and multi-shell wave functions are similar. In fact, the $^{40}$Ca core in $^{41}$Sc is broken by the same amount as in $^{41}$Ca, which shows that the robustness of the shell closure cannot be solely concluded from the behavior of the magnetic moments.

\begin{figure}[t]
    \centering
    \includegraphics[clip,width=\columnwidth]{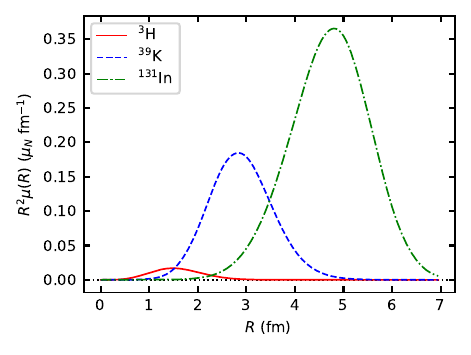}
    \caption{Contributions from the Sachs term for the single proton-hole systems $^{3}$H, $^{39}$K, and $^{131}$In as a function of the center-of-mass position of the two nucleons $R$ relative to the center of the nucleus. The results for $^{39}$K and $^{131}$In are obtained with VS-IMSRG(2), while $^{3}$H is computed with the no-core shell model~\cite{Seutin2023} at $N_{\rm max}=20$ and $\hbar\omega = 16$\,MeV.}
    \label{fig:Rdep}
\end{figure}

Analyzing the two contributions from 2BC for the odd-mass nuclei studied in Fig.~\ref{fig:mu_1p_1h}, we observe that the Sachs contribution, Eq.~\eqref{eq:mu_Sachs}, becomes larger with increasing mass number.
For example, the ratio $|\bra \mu^{\rm Sachs}\ket / \bra \mu^{\rm intr} \ket|$ for the ground state is about $0.1$ for $^{3}$H, while it is $10$ for $^{131}$In.
This can be understood by taking the simple single-particle limit, i.e., approximating the ground state by closed-shell core with the last unpaired nucleon occupying a single-particle orbit with collective index $p$. In this limit, the 2BC contribution to the magnetic moment is given by
\begin{equation}
\bra \mu_{\rm 2B} \ket \approx \sum_{q \in {\rm Core}} \bra pq | \tilde{\mu}_{\rm 2B} | pq \ket \,, 
\end{equation}
where $\bra pq |\tilde{\mu}_{\rm 2B} | pq \ket$ is a two-body matrix element of $\mu_{\rm 2B}$ with the appropriate angular momentum coupling factors.
The matrix element $\bra pq |\tilde{\mu}_{\rm 2B} |pq\ket$ can only be significant if the single-particle states $p$ and $q$ have a large overlap, because the 2BC contribution is of pion range. Therefore, if the nucleon is in a high orbital angular momentum state $p$ located near the surface of the nucleus (e.g., with $l=4$ in $^{131}$In), the center-of-mass coordinate of the two nucleons in orbits $p$ and $q$ should also be near the surface at large $R$. Since the Sachs term is proportional to the center-of-mass coordinate $R$, this explains that its expectation value grows with increasing $A$.

To test this picture, we have calculated the Sachs term distribution $\mu(R)$ as a function of the center-of-mass position of the two nucleons $R$ relative to the center of the nucleus, which can be computed as
\begin{equation}
 \mu(R) = \left\bra \sum_{i<j} \mu^{\rm Sachs}_{ij}\frac{1}{R^{2}}\delta(R - R_{ij}) \right\ket \,.
\end{equation}
Note that one obtains the Sachs 2BC contribution after integrating over $R$,
\begin{equation}
\bra \mu^{\rm Sachs}\ket = \int dRR^{2}\mu(R) \,.
\end{equation}
As shown in Fig.~\ref{fig:Rdep} and expected based on the arguments given above, the maximum of the Sachs term distribution moves to a larger $R$ as the mass number increases. Since the Sachs term is dominant for medium-mass nuclei and tends to grow towards heavier systems, the inclusion of the 2BC contribution is critical to reproduce magnetic dipole moments in heavy-mass nuclei.

\begin{figure}[t]
    \centering
    \includegraphics[clip,width=\columnwidth]{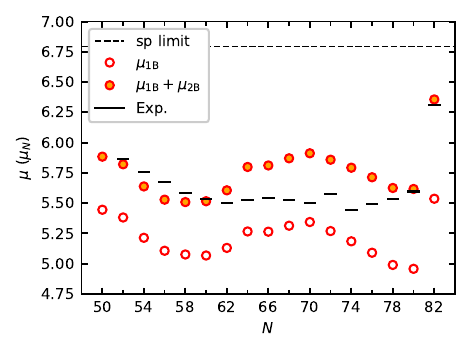}
    \caption{Magnetic dipole moments of the $9/2^{+}$ ground state for the odd-mass indium isotopes computed with the VS-IMSRG(2) including 2BC, in comparison to experiment~\cite{Vernon2022,Karthein2023}.}
    \label{fig:mu_In}
\end{figure}

Finally, we show the magnetic dipole moments of the $9/2^{+}$ ground state for the indium isotopes in Fig.~\ref{fig:mu_In}, where previous VS-IMSRG calculations with $\mu_{\rm 1B}$ underestimated the magnetic moments~\cite{Vernon2022}.
The calculational setup over the isotopic chain is the same as for $^{131}$In in Fig.~\ref{fig:mu_1p_1h} (for details see Supplemental Material).
We also computed the magnetic moments with the $\Delta {\rm N}^{2}{\rm LO}_{\rm GO}(394)$ interaction~\cite{Jiang2020} and observed that the interaction dependence is about a few percent, averaging over the isotopes, which is significantly smaller than the 2BC contribution. The results with $\Delta {\rm N}^{2}{\rm LO}_{\rm GO}(394)$ can be found in Supplemental Material.
The 2BC contributions $\mu_{\rm 2B}$ systematically increase the results towards experimental values, and the sudden increase to the shell closure at $N=82$ is excellently reproduced with 2BC included.
The nearly constant 2BC contribution may be attributed to the dominance of Sachs term in heavier nuclei. Since the increase of nuclear radii along the studied indium chain is weak, the Sachs term contribution approximately remains constant.
However, the detailed behavior around $N=70$ is not satisfactory, and a more sophisticated many-body treatment, through an explicit inclusion of deformation, for these mid-shell isotopes would be needed~\cite{Vernon2022}.
Moreover, the behavior of magnetic moments towards $N=50$ is intriguing.
Naively, the $N=50$ magic number should show similar jump as for $N=82$, which is also found in results from density-functional theory~\cite{Vernon2022}.
In our calculations, the behavior towards $N=50$ is already different at the $\mu_{1B}$ level, where magnetic moments increase smoothly with relatively large $N=52$ and $N=54$ results. This is because the single proton-hole configuration $|(\pi g_{9/2})^{-1}\ket$ is more pronounced in these isotopes.
This favored spherical structure at $N=52$ and $54$ may, however, be due to the difficulty of the VS-IMSRG to capture deformation, that would arise from the near degeneracy of the neutron $g_{7/2}$ and $d_{5/2}$ orbitals.
Exploring this requires further experiments and theoretical calculations with methods based on deformed reference states (see, e.g., Ref.~\cite{Hagen2022}).

In summary, we have explored the impact of the leading 2BC on magnetic dipole moments from medium-mass to heavy nuclei using the ab initio VS-IMSRG.
The 2BC contributions globally improve the agreement with experimental magnetic moments of near doubly magic nuclei from oxygen to bismuth. For the case of $^{41}$Ca, in addition to 2BC, we found that multi-shell effects are important, due to the weaker closed-shell nature of the $^{40}$Ca core.
Moreover, we have found that the 2BC contributions increase in heavier systems. This could be understood by the structure of the center-of-mass dependent Sachs term, which is enhanced near the nuclear surface. Finally, including 2BC leads to an excellent reproduction of the magnetic moments of the indium isotopes near $N=82$ and around $N=60$. Further work is needed to better include deformation effects for heavy open-shell nuclei and to shed light on the evolution towards $N=50$ in indium isotopes.
Our work shows that the inclusion of 2BC for the exploration of EM observables is a frontier. This opens up exciting opportunities after our first global survey of magnetic moments, exploring other EM properties of nuclei and including higher-order 2BC consistently in chiral EFT, enabling us to perform a full uncertainty quantification and detailed comparisons with experimental data.

\begin{acknowledgments}
We thank B.~Acharya, J.~Dobaczewski, G.~Hagen, B.~S.~Hu, T.~Papenbrock, and S.~R.~Stroberg for useful discussions. 
For the VS-IMSRG and subsequent configuration-interaction calculations, \texttt{imsrg++}~\cite{imsrg++} and \texttt{KSHELL}~\cite{KSHELL} codes were used.
This work was supported in part by the European Research Council (ERC) under the European Union’s Horizon 2020 research and innovation programme (Grant Agreement No. 101020842), the Deutsche Forschungsgemeinschaft (DFG, German Research Foundation) -- Project-ID 279384907 -- SFB 1245, NSERC under grants SAPIN-2018-00027 and RGPAS-2018-522453, and Office of Nuclear Physics, U.S. Department of Energy, under grant DE-SC0021176.
TRIUMF receives funding via a contribution through the National Research Council of Canada.
Computations were performed with an allocation of computing resources at the J\"ulich Supercomputing Center and on Cedar at WestGrid and Compute Canada.
\end{acknowledgments}

\bibliography{library}

\newpage

\section{\label{Supplementary}Supplemental Material}

\begin{figure}[t]
    \centering
    \includegraphics[clip,width=\columnwidth]{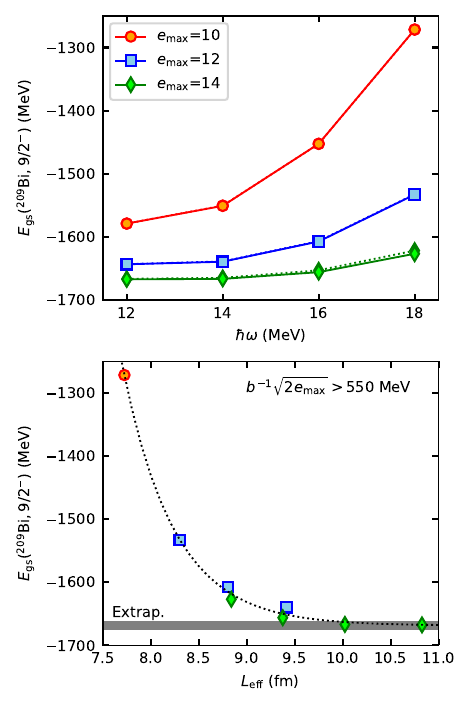}
    \caption{Convergence of the ground-state energy of $^{209}$Bi computed with the VS-IMSRG(2). In the top panel, the symbols connected with the dotted lines are the results with $E_{\rm 3max}=28$, while those connected with the solid lines are $E_{\rm 3max}$ extrapolated values. For $e_{\rm max}=10, 12$ the dotted lines are behind the solid lines.
    The bottom panel shows the extrapolation based on Ref.~\cite{Furnstahl2015}, and the gray band indicates the extrapolated energy.}
    \label{fig:conv_Bi209}
\end{figure}

\begin{figure}[t]
    \centering
    \includegraphics[clip,width=\columnwidth]{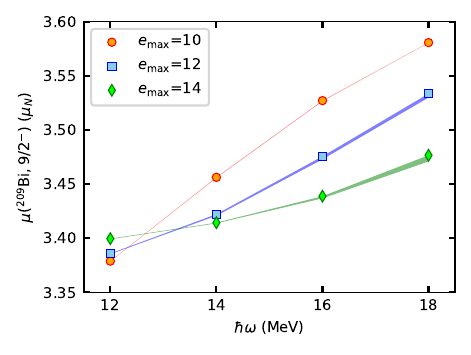}
    \caption{Convergence of the ground-state magnetic dipole moment including 2BC of $^{209}$Bi computed with the VS-IMSRG(2).}
    \label{fig:Bi209_mu_conv}
\end{figure}

\subsection{Calculational setup}

The calculations are performed with the NN+3N Hamiltonian expressed in spherical harmonic oscillator (HO) basis states with an $e_{\rm max}$ truncation defined by $e_{\rm max} = \max(2n+l)$, where $n$ and $l$ are the radial and angular momentum quantum numbers of the single-particle state.
For 3N matrix elements, we introduce a further truncation to make calculations possible with $E_{\rm 3max}$ defined by the sum of three-body HO quanta $E_{\rm 3max}=\max(2n_{1}+l_{1}+2n_{2}+l_{2}+2n_{3}+l_{3})$. Note that very recently, calculations without $E_{\rm 3max}$ truncation, i.e., $E_{\rm 3max}=3e_{\rm max}$, became feasible~\cite{Hebeler2023} by using the normal-ordered two-body approximation within the Jacobi basis.

For $^{17}$O, $^{17}$F, $^{37}$Cl, $^{39}$K, $^{37}$Ca, $^{39}$Ca, $^{41}$Ca, and $^{41}$Sc, the calculations were performed with $e_{\rm max}=12$ and $E_{\rm 3max}=24$.
For $^{131}$In and $^{133}$Sb, $e_{\rm max}=14$ and $E_{\rm 3max}=24$ were used.
For the heaviest systems $^{207}$Tl and $^{208}$Bi in this work, we used $e_{\rm max}=14$ and $E_{\rm 3max}=28$ are used.
As for the valence space, we used the standard $sd$-shell valence space for $^{17}$O, $^{17}$F, $^{37}$Cl, $^{39}$K, $^{37}$Ca, and $^{39}$Ca.
The $pf$-shell space was used for $^{41}$Ca and $^{41}$Sc.
The indium isotopes were computed with the proton $\{1p_{1/2}, 1p_{3/2}, 0f_{5/2}, 0g_{9/2}\}$ and neutron $\{2s_{1/2}, 1d_{3/2}, 1d_{5/2}, 0g_{7/2}, 0h_{11/2}\}$ above the $^{78}$Ni core.
The $^{133}$Sb calculation was performed with proton-neutron $\{2s_{1/2}, 1d_{3/2}, 1d_{5/2}, 0g_{7/2}, 0h_{11/2}\}$ valence space above the $^{100}$Sn core.
The proton $\{2s_{1/2}, 1d_{3/2}, 1d_{5/2}, 0g_{7/2}, 0h_{11/2}\}$ above the $^{176}$Sn core and $\{2p_{1/2}, 2p_{3/2}, 1f_{5/2}, 1f_{7/2}, 0h_{9/2}, 0i_{13/2}\}$ above the $^{208}$Pb core are chosen for $^{207}$Tl and $^{209}$Bi, respectively.
The calculated results are tabulated in Table~\ref{tab:res}.
For the multi-shell calculations given in Fig.~\ref{fig:mu_Ca}, the proton-neutron $\{1s_{1/2}, 0d_{3/2}, 1p_{3/2}, 0f_{7/2}\}$ valence-space above $^{28}$Si was used with the modified IMSRG generator~\cite{Miyagi2020}. 
Note that we observed that the center-of-mass motion is well decoupled.

\begin{table*}[t]
\caption{\label{tab:res} Results for the magnetic dipole moments shown in Fig.~\ref{fig:mu_1p_1h} with the relevant details of the VS-IMSRG calculations.}
\centering
\begin{tabular}{l|l|r cl|c| rrr}
\hline\hline
Nucleus \: & Valence space & $e_{\rm max}$ & $E_{\rm 3max}$ & $\hbar\omega$ (MeV) \: & State \: & $\mu_{\rm 1B}$ ($\mu_{N}$) & $\mu_{\rm 2B}$ ($\mu_{N}$) & $\mu_{\rm 1B}+\mu_{\rm 2B}$ ($\mu_{N}$) \\
\hline
    $^{17}$O &                                           $sd$-shell &     12 &     24 &     16 &   $5/2^{+}_{1}$ &   $-1.752$ &   $-0.161$ &   $-1.913$ \\
    $^{17}$F &                                           $sd$-shell &     12 &     24 &     16 &   $5/2^{+}_{1}$ &    4.626 &    0.157 &    4.783 \\
   $^{37}$Cl &                                           $sd$-shell &     12 &     24 &     16 &   $3/2^{+}_{1}$ &    0.290 &    0.290 &    0.579 \\
    $^{39}$K &                                           $sd$-shell &     12 &     24 &     16 &   $3/2^{+}_{1}$ &   $-0.035$ &    0.274 &    0.240 \\
   $^{37}$Ca &                                           $sd$-shell &     12 &     24 &     16 &   $3/2^{+}_{1}$ &    1.055 &   $-0.299$ &    0.756 \\
   $^{39}$Ca &                                           $sd$-shell &     12 &     24 &     16 &   $3/2^{+}_{1}$ &    1.349 &   $-0.284$ &    1.065 \\
   $^{41}$Ca &                                           $pf$-shell &     12 &     24 &     16 &   $7/2^{-}_{1}$ &   $-1.610$ &   $-0.313$ &   $-1.923$ \\
   $^{41}$Sc &                                           $pf$-shell &     12 &     24 &     16 &   $7/2^{-}_{1}$ &    5.482 &    0.308 &    5.790 \\
  $^{131}$In &             $28 \leq Z \leq 50, 50 \leq N \leq 82$ &     14 &     24 &     16 &   $9/2^{+}_{1}$ &    5.536 &    0.820 &    6.357 \\
  $^{133}$Sb &                              $50 \leq Z,N \leq 82$ &     14 &     24 &     12 &   $7/2^{+}_{1}$ &    1.915 &    0.545 &    2.460 \\
  $^{207}$Tl &                         $50 \leq Z \leq 82, N=126$ &     14 &     28 &     12 &   $1/2^{+}_{1}$ &    2.212 &    0.091 &    2.303 \\
  $^{209}$Bi &                        $82 \leq Z \leq 126, N=126$ &     14 &     28 &     12 &   $9/2^{-}_{1}$ &    2.782 &    0.617 &    3.399 \\
\hline\hline
\end{tabular}
\end{table*}

\subsection{Convergence of $^{209}$Bi}

Figure~\ref{fig:conv_Bi209} shows the convergence of the calculations for the heaviest nucleus studied in this work, $^{209}$Bi.
In the top panel, we show the HO basis frequency dependence and the impact of the $E_{\rm 3max}$ extrapolation~\cite{Miyagi2022} illustrated by the difference between dotted and solid lines.
In the bottom panel, the infrared scale extrapolation towards infinite model space is shown.
As detailed in Ref.~\cite{Furnstahl2015}, the energies can be given as a function of $L_{\rm eff}$ rather than $e_{\rm max}$ and $\hbar\omega$ and extrapolated with $E(L_{\rm eff}) = E_{\infty} + A_{\infty} \exp(-2k_{\infty}L_{\rm eff})$, where $E_{\infty}$ is the converged energy.
The length scale $L_{\rm eff}$ is defined as
\begin{equation}
L_{\rm eff} = \left(\frac{\sum_{nl} n^{\rm occ}_{nl}a^{2}_{nl}}{\sum_{nl} n^{\rm occ}_{nl}\kappa^{2}_{nl}}\right)^{1/2} \,,
\end{equation}
with the occupation number $n^{\rm occ}_{nl}$ of an orbit specified by $n$ and $l$, the $(n+1)$-th zero of the spherical Bessel function $a_{nl}$, and the eigenvalue of the single-particle squared momentum operator $\kappa^{2}_{nl}$.
Following Ref.~\cite{Furnstahl2015}, $\kappa^{2}_{nl}$ is approximated as
\begin{equation}
\kappa^{2}_{nl} \approx \frac{a^{2}_{nl}}{2(N_{l}+3/2 + 2)b^{2}} \,,
\end{equation}
where $N_{l}$ is equal to $e_{\rm max}$ [$e_{\rm max}-1$] if $e_{\rm max} + l = 0 \pmod{2}$ [$e_{\rm max} + l = 1 \pmod{2}$].
Note that each $(e_{\rm max}, \hbar\omega)$ point also has a corresponding ultraviolet scale $\Lambda_{\rm UV} \sim b^{-1}\sqrt{2e_{\rm max}}$ with oscillator length parameter $b= \sqrt{\hbar / m\omega}$, and in principle we should extrapolate with respect to $\Lambda_{\rm UV}$ as well.
However, the functional form for $\Lambda_{\rm UV}$ is not as clear as $L_{\rm eff}$, and here we only adopt the points having large enough $\Lambda_{\rm UV}$ compared to the scale of the employed interaction, $\sim 400$ MeV in the case of the 1.8/2.0 (EM) interaction.
In the bottom panel, we show the extrapolation with the points satisfying $b^{-1}\sqrt{2e_{\rm max}} > \Lambda_{\rm UV}$ with $\Lambda_{\rm UV} = 550$ MeV.
We observed that the extrapolated energies are somewhat dependent on $\Lambda_{\rm UV}$, and thus we combined $\Lambda_{\rm UV}=550$ and $600$ MeV results using the weights $w_{\Lambda_{\rm UV}} = N / \chi_{\Lambda_{\rm UV}}^{2}$ with the normalization $N$ and $\chi^{2}_{\rm UV}$ of the fitting procedure.
In the end, we estimate that the extrapolation yields $-4\pm 12$ MeV with respect to the calculated ground-state energy at $e_{\rm max}=14$, $E_{\rm 3max}=28$, and $\hbar\omega=12$ MeV. 
We expect a similar convergence pattern for $^{207}$Tl and apply the same shift and uncertainty for the $^{207}$Tl ground-state energy.

In Fig.~\ref{fig:Bi209_mu_conv}, the convergence pattern of the magnetic dipole moment is shown for the $9/2^{-}$ ground state of $^{209}$Bi, the heaviest case in this work.
The bands connecting symbols are obtained from $E_{\rm 3max}=26$ and $28$ results.
As the band width is smaller than the $e_{\rm max}$ variation, the $E_{\rm 3max}$ dependence is negligible.
Unlike the ground-state energy, extrapolation formulas for the magnetic dipole moment are unknown, and one needs to compute it in a sufficiently large model space.
As seen in Fig.~\ref{fig:Bi209_mu_conv}, the $\hbar\omega$ dependence becomes weaker with increasing $e_{\rm max}$, and the result at $e_{\rm max}=14$ and around $\hbar\omega=14$ MeV is expected to be the converged one.
Therefore the result at $e_{\rm max}=14$ and $\hbar\omega=12$ MeV is converged within less than 1\% level, which is less than the typical truncation errors in chiral EFT and the many-body VS-IMSRG calculation.

\subsection{Magnetic moments of indium isotopes with $\Delta{\rm N}^{2}{\rm LO}_{\rm GO}(394)$ interaction}

\begin{figure}[t]
    \centering
    \includegraphics[clip,width=\columnwidth]{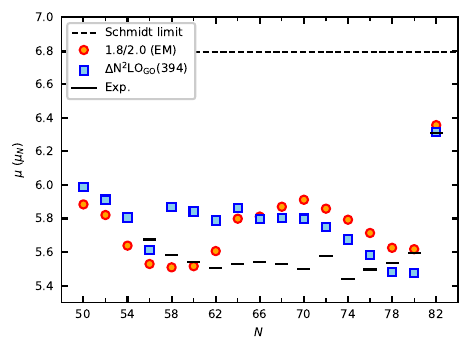}
    \caption{Magnetic dipole moments of the $9/2^{+}$ ground state for the odd-mass indium isotopes calculated from the 1.8/2.0 (EM)~\cite{Hebeler2011} and $\Delta{\rm N}^{2}{\rm LO}_{\rm GO}(394)$ interactions with the 2BC contributions, computed at the VS-IMSRG(2) level.}
    \label{fig:mu_In_intdep}
\end{figure}

Figure~\ref{fig:mu_In_intdep} shows the magnetic dipole results with $\mu_{\rm 1B} + \mu_{\rm 2B}$ for odd-mass indium isotopes calculated from the $\Delta{\rm N}^{2}{\rm LO}_{\rm GO}(394)$~\cite{Jiang2020} interaction, in addition to the 1.8/2.0 (EM)~\cite{Hebeler2011} results shown in the main text.

\end{document}